\documentclass[aps,preprint,prd,nofootinbib]{revtex4}
\usepackage{graphicx}
\usepackage{psfrag}

\usepackage{subfigure}
\usepackage{array}
\usepackage{graphicx}
\usepackage{amsmath}
\usepackage{amssymb}
\usepackage{mathrsfs}
\usepackage{bm}
\usepackage{slashed}
\usepackage{threeparttable}
\usepackage{multirow}
\usepackage[colorlinks, linkcolor=black, anchorcolor=black, citecolor=black]{hyperref}
\usepackage[amsmath,thmmarks,hyperref]{ntheorem}
\usepackage{appendix}
\usepackage{soul}

\allowdisplaybreaks[4]

\begin{document}

\title{Semi-leptonic decays of $B^\ast$, $B_s^\ast$, and $B_c^\ast$ with the Bethe-Salpeter method}
\author{Tianhong Wang\footnote{thwang@hit.edu.cn}, Yue Jiang, Tian Zhou, Xiao-Ze Tan, and~Guo-Li Wang\footnote{gl\_wang@hit.edu.cn}\\}
\address{Department of Physics, Harbin Institute of Technology, Harbin, 150001, China}

\baselineskip=20pt

\begin{abstract}
In this paper we study the semileptonic decays of $B^\ast$, $B_s^\ast$, and $B_c^\ast$ by using the Bethe-Salpeter method with instantaneous approximation. Both the $V\to Pl^-\bar\nu_l$ and $V\to Vl^-\bar\nu_l$ cases are considered. 
The largest partial width of these channels is of the order of $10^{-14}$ GeV. The branching ratios of these semileptonic decays are also estimated by using the partial width of the one photon decay channel to approximate the total width. For $B^{\ast-}\to D^{(\ast)0}e^-\bar\nu_e$ and $B_s^{\ast0}\to D_s^{(\ast)+}e^-\bar\nu_e$, the branching ratios are of the order of $10^{-8}$ and $10^{-7}$, respectively. For $B_c^{\ast-}$, the $J/\psi e^-\bar\nu_e$ and $B_s^{\ast0}e^-\bar\nu_e$ channels have the largest branching ratio, which is of the order of $10^{-6}$.

\end{abstract}

\maketitle

\section{Introduction}
The $b$-flavored pseudoscalar heavy mesons, namely $B$, $B_s$, and $B_c$ have been studied extensively both in theories and experiments. The most important reason for this is that they can only decay weakly, hence provides an opportunity to do the precision investigation tests on the Standard Model (SM). However, the vector mesons $B^\ast$, $B_s^\ast$, and $B_c^\ast$ still lack enough experimental results (see Ref.~\cite{pdg}), because both their production rates and detection efficiency are lower than the pseudoscalar partners. This situation will change as LHCb collecting more and more data, which makes their precise detection be possible. For example, when LHC runs at 14 TeV, the cross section for the hadronic production of $B_c^\ast$ is predicted to be 33.1 nb~\cite{chang05}. If the integrated luminosity is taken to be 1 ${\rm fb}^{-1}$, there are about $10^7$ $B_c^\ast$ expected per year. Also, the future B-factories, such as Belle II, will also provide more information for these particles. So the theoretical studies of these vector heavy-light mesons becomes more and more necessary.

A notable property for these particles is that their masses are not large enough to decay to the corresponding pseudoscalar partner and a light meson, such as $\pi$, $K$, {\sl et al.}. So these particles cannot decay strongly, but can only decay weakly and electromagnetically. As a result, the partial widths of the electromagnetic decay channels, especially the one-photon decay channel, are dominant, which can be used to estimate the total widths. Theoretically, these channels have partial widths less than 1 keV~\cite{cheu14,choi07,pat17}. This makes the branching ratios of  their weak decay models may be within the detection ability of current experiments.

Recently, there are some interests of finding new physics in the $B_{d,s}^\ast$ meson decays~\cite{kho15,Grin16,sahoo17,Kumar18}, such as the $B_s^\ast\to\mu^+\mu^-$ channel, which has the branching ratio around $10^{-12}$ in the SM~\cite{xu16}. This result is too small to be detected nowadays at the LHC (although there is possibility by the end of run III of the LHC as Ref.~\cite{Grin16} mentioned). However, the smileptonic channels could have larger branching fractions so that they can be investigated experimentally.  This is the case for their pseudoscalar partners $B_{d,s}$, of which the $l^+l^-$ decays have branching fractions much smaller than those of the semilptonic decay channels~\cite{pdg}. 

Until now, there are only limited theoretical calculations of such decay channels carried out. In Ref.~\cite{qc01}, the smileptonic decays of the $B^\ast_{d,s}$ with a final pseudoscalar meson were studied. In their work, the hadronic transition matrix elements are calculated in the Bauer-Stech-Wirbel (BSW) model. However, using a different method to study such channels are necessary, as by comparing the results of different models can make us to know how large they are model dependent. In this paper, we will use the instantaneously approximated Bethe-Salpeter (BS) method which also has been applied extensively to deal with weak decays of $B_{q}$ mesons~\cite{zhang10,fu11,li17}. As the instantaneous approximation is reliable only for the heavy mesons, we will focus on the decay channels with the final meson also being heavy. There are also some approaches to deal with light mesons, such as the Dyson-Schwinger equation (DSE) model~\cite{IvanPLB, IvanPRD}. Both in the DSE model and our model, the calculation of the transition amplitude contains two main elements, the quark propagator and the meson amplitude. In the DSE model, the dressed-quark propagator is applied, where the effects of confinement and the dynamical chiral symmetry breaking are considered, which are more related to QCD; for the heavy meson amplitude, usually a simple function, such as the exponential function is assumed. In our model, a simple form of the quark propagator is applied, and by solving the instantaneous BS equations, we can get the wave function of the heavy meson, which can be used directly to calculate the form factors. Besides the $B^\ast_{d,s}$ mesons, we will also study the semileptonic decays of the $B^\ast_c$ meson, which have been studied by even limited work. For instance, in Ref.~\cite{wzg14}, the QCD sum rules approach is applied to study its semileptonic decays, but only the $B_c^{\ast-}\to\eta_cl^-\bar\nu_l$ channels are considered. One reason for this may be that this particle has not been found in experiments. However, LHCb have made some efforts very recently to find excited $B_c$ states~\cite{LHCb}. We expect that the $B_c^\ast$ state can be found in the near future. So the study of its decay properties is also of interest. 

The article is organized as follows. In Section II we give the theoretical formalism of the calculation. The hadronic transition amplitudes both for the $V\to P$ and $V\to V$ processes are presented. The numerical results of the partial widths, the branching fractions, the leptonic spectra, and corresponding discussions are given in Section III. Finally, we conclude in Section IV.

\section{Theoretical Formalism}

The wave function $\chi_{_P}(q)$ of the two-body bound state fulfills the BS equation
\begin{equation}\label{BS}
\begin{aligned}
S^{-1}_1(p_1)\chi_{_P}(q)S^{-1}_2(-p_2)=i\int\frac{d^4k}{(2\pi)^4} V(P; q,k)\chi_{_P}(k),
\end{aligned}
\end{equation}
where $p_1$ and $p_2$ are the momenta of quark and antiquark, respectively; $S_1(p_1)$ and $S_2(-p_2)$ are propagators of quark and antiquark, respectively; $P$ is the momentum the bound state; $q$ is the relative momentum between quark and antiquark; $V(P; q,k)$ is the interaction kernel. By taking the  instantaneous approximation $V(P; q, k) \approx V(P; q_\perp, k_\perp)$ and defining $\varphi_{_P}(q_\perp)\equiv i\int\frac{dq^0}{2\pi}\chi_{_P}(q)$, the BS equation can be reduced to the Salpeter equation~\cite{Kim}
\begin{equation}
\label{salpeter}
\begin{aligned}
&(M-\omega_1-\omega_2)\varphi_{_P}^{++}(q_\perp) = \Lambda_1^+\eta_{_P}(q_\perp)\Lambda_2^+,\\
&(M+\omega_1+\omega_2)\varphi_{_P}^{--}(q_\perp) = - \Lambda_1^-\eta_{_P}(q_\perp)\Lambda_2^-,\\ &\varphi_{_P}^{+-}(q_\perp) = \varphi_{_P}^{-+}(q_\perp) = 0,
\end{aligned}
\end{equation}
where $q^\mu_\perp=q^\mu-\frac{P\cdot q}{M^2}P^\mu$, $\omega_1=\sqrt{m_1^2-q_\perp^2}$, and $\omega_2=\sqrt{m_2^2-q_\perp^2}$; $m_1$ and $m_2$ are the masses of quark and antiquark, respectively.  In the above equation, we have defined 
\begin{equation}
\label{eta}
\eta_{_P}(q_\perp)=\int\frac{d^3k_\perp}{(2\pi)^3} V(P; q_\perp,k_\perp)\varphi_{_P}(k_\perp),
\end{equation}
and 
\begin{equation}
\varphi_{_P}^{\pm\pm}(q_\perp)=\Lambda^{\pm}_1\frac{\slashed P}{M}\varphi_{_P}(q_\perp)\frac{\slashed P}{M}\Lambda^{\pm}_2,
\end{equation} 
where $\Lambda_i^{\pm}=\frac{1}{2\omega_i}\left[\frac{\slashed P}{M}\omega_i\mp(-1)^{i} (\slashed q_\perp + m_i)\right]$ is the projection operator. The expressions for $\varphi$ and $\varphi^{++}$ are given in the Appendix.

We use the Cornell-like interaction potential, which in the momentum space has the form~\cite{Kim}
\begin{equation}
\begin{aligned}
\label{Cornell}
V({\vec q})=V_s(\vec{q})
+\gamma_0\otimes\gamma^0V_v(\vec{q}),
\end{aligned}
\end{equation}
where
\begin{equation}
\begin{aligned}
&V_{s}(\vec{q})
=-\left(\frac{\lambda}{\alpha}+V_0\right)\delta^{3}(\vec{q})
+\frac{\lambda}{\pi^{2}}\frac{1}{(\vec{q}^{2}+\alpha^{2})^{2}},\\
&V_v(\vec{q})=-\frac{2}{3\pi^{2}}
\frac{\alpha_{s}(\vec{q})}{\vec{q}^{2}+\alpha^{2}},\\
&\alpha_s(\vec{q})=\frac{12\pi}{27}
\frac{1}{{\rm{ln}}\left(a+\frac{\vec{q}^2}{\Lambda^2_{QCD}}\right)}.
\end{aligned}
\end{equation}
The parameters involved are $a=e=2.71828$, $\alpha=0.06$ GeV, $\lambda=0.21$ ${\rm GeV}^2$, $\Lambda_{QCD}=0.27$ GeV, $m_b=4.96$ GeV, $m_c=1.62$ GeV, $m_s=0.5$ GeV, $m_u=0.305$ GeV, $m_d=0.311$ GeV; $V_0$ is decided by fitting the mass of the ground state.

The Feynman diagram for the semileptonic decay is presented in Figure 1. The amplitude of this process can be written as the product of the leptonic part and the hadronic transition matrix element
\begin{equation}
  \mathcal{M}=\frac{G_{F}}{\sqrt{2}}V_{Qq}\bar{u}_{l}\gamma_{\mu}(1-\gamma_{5})v_{\bar{\nu}_{l}}\langle P_f|J^{\mu}|P, \epsilon\rangle,
\end{equation}
where $G_F$ is the Fermi constant; $V_{Qq}$ is the Cabibbo-Kobayashi-Maskawa (CKM) matrix element; $P$ and $P_f$ are the momenta of the initial and final meson, respectively; $\epsilon$ is the polarization vector of the initial meson; $J^\mu=\bar q\gamma^\mu(1-\gamma^5)Q=V^\mu-A^\mu$ is the weak current. 

\begin{figure}
\centering
\includegraphics[width=0.48\textwidth]{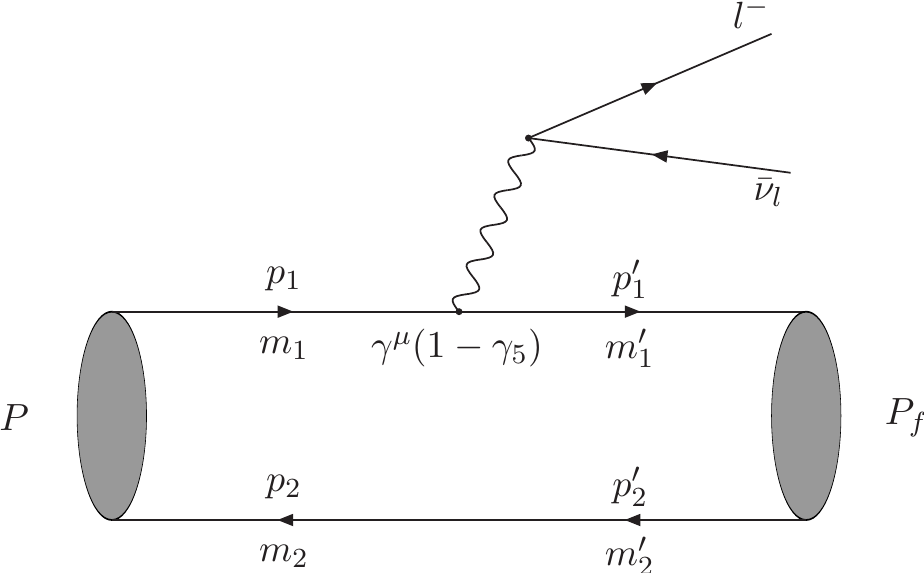}
\caption[]{The Feynman diagram of the semileptonic decay of the vector meson.}
\end{figure}

Within Mandelstam formalism, the hadronic transition matrix element can be written as the overlap integral of the instantaneous BS wave functions of the initial and final heavy mesons~\cite{chang01},
\begin{equation}
\begin{aligned}
\langle P_f|J^{\mu}|P, \epsilon\rangle=\int\frac{d{\vec{q}}}{(2\pi)^{3}}\textrm{Tr}\left[\frac{\rlap{$/$}P}{M}\overline{\varphi_{_{P_f}}^{++}}({\vec{q}_f})\gamma_{\mu}(1-\gamma_{5})\varphi_{_{P}}^{++}({\vec{q}})\right],
\end{aligned}
\end{equation}
where $\vec q$ and $\vec q_f$ are the relative three-momenta between quark and antiquark within the initial and final mesons, respectively; $\varphi_{_{P}}^{++}({\vec{q}})$ and $\varphi_{_{P_f}}^{++}({\vec{q_f}})$ are the positive energy parts of the wave functions of initial and final mesons, respectively, whose explicit expressions can be found in the Appendix. The final meson can be a pseudoscalar or a vector, and we give the expressions of hadronic transition matrix elements for both cases.

For the $1^{-}\rightarrow 0^-$ channel~\cite{wang17}
\begin{equation}
\begin{aligned}
&\langle P_f|V^{\mu}|P, \epsilon\rangle=\frac{2s_1}{M+M_f}i\epsilon^{\mu\nu\rho\sigma}\epsilon_\nu P_\rho P_{f\sigma},\\
&\langle P_f|A^{\mu}|P, \epsilon\rangle= s_2(M+M_f)\epsilon^\mu - (s_3P^\mu - s_4P_f^\mu)\frac{\epsilon\cdot P_f}{M},
\end{aligned}
\end{equation}
where $M$ and $M_f$ are the masses of the initial and final mesons, respectively; $s_1\sim s_4$ are form factors which are the integrals of $\vec q$.

For the $1^{-}\rightarrow 1^-$ channel~\cite{wang17}
\begin{equation}
\begin{aligned}
&\langle P_f, \epsilon_f|V^{\mu}|P, \epsilon\rangle=(t_1P^\mu + t_2P_f^\mu)\frac{\epsilon\cdot P_f\epsilon_f\cdot P}{M^2}-t_3\epsilon^\mu\epsilon_f\cdot P - t_4\epsilon_f^\mu\epsilon\cdot P_f  \\
&~~~~~~~~~~~~~~~~~~~~~~~+ (t_5P^\mu + t_6P_f^\mu)\epsilon\cdot\epsilon_f,\\
&\langle P_f, \epsilon_f|A^{\mu}|P, \epsilon\rangle= i\epsilon^{\mu\alpha\gamma\delta}\frac{P_{\gamma} P_{f\delta} }{M^2}(h_1\epsilon_\alpha\epsilon_f\cdot P + h_2\epsilon_{f\alpha}\epsilon\cdot P_f) + i\epsilon^{\mu\alpha\beta\gamma}\epsilon_{\alpha}\epsilon_{f\beta}\\
&~~~~~~~~~~~~~~~~~~~~~~~ \times(h_3P_\gamma+ h_4P_{f\gamma}),
\end{aligned}
\end{equation}
where $\epsilon_f$ is the polarization vector of the final meson; $t_1\sim t_6$ and $h_1\sim h_4$ are the form factors.

The partial decay width is achieved by finishing the phase space integral
\begin{equation}
\begin{aligned}
\varGamma=\frac{1}{3}\frac{1}{8M(2\pi)^3}\int dE_{l}dE_{f}\sum_{\lambda}|\mathcal{M}|^{2},
\end{aligned}
\end{equation}
where $E_l$ and $E_f$ are the energy of charged lepton and final meson, respectively; $\lambda$ represents the polarization indexes of both initial and final mesons. From this, one can also easily calculate the differential partial widths.

\section{Results and Discussions}

The $B^{\ast-}$ and $B_s^{\ast0}$ mesons have been found experimentally~\cite{pdg} with masses $M(B^{\ast-})=5.325$ GeV and $M(B_s^{\ast0})=5.415$ GeV, respectively. However, there is still not enough experimental data about both their total and partial widths. As the strong decays are forbidden by the phase space, the total decay widths of these vector $b$-flavored mesons can be estimated by the partial width of the single-photon decay channel~\cite{cheu14, choi07}
\begin{equation}
\begin{aligned}
&\Gamma_{B^{\ast-}}\simeq \Gamma(B^{\ast+}\to B^+\gamma) = 468^{+73}_{-75}~ {\rm eV},\\
&\Gamma_{B_s^{\ast0}}\simeq \Gamma(B_s^{\ast0}\to B_s^0\gamma) = 68\pm17~ {\rm eV}.
\end{aligned}
\end{equation}
The $B_c^{\ast-}$ meson has not been found experimentally. Here we take the value $M(B_c^{\ast-})=$ 6.333 GeV predicted by the quark petential model~\cite{ebert11}. The one photon decay width is calculated recently in Ref.~\cite{pat17}, which can be used to approximate the total width 
\begin{equation}
\Gamma_{B_c^{\ast-}}\simeq \Gamma(B_c^{\ast-}\to B_c^-\gamma) = 23~ {\rm eV}.
\end{equation}
One notices that the one photon decay widths of $B_s^{\ast0}$ and $B_c^{\ast-}$ are about one order smaller than that of the $B^{\ast-}$ meson. These results surely are model dependent, however, the order of magnitude should be affirmatory.

The partial widths of the $V\to P$ channels are presented in Table I. All the cases when $l^-=e^-,~\mu^-$, and $\tau^-$ are considered. For $B^{\ast-}$ and $B_s^{\ast0}$, the decay channels with the same charged lepton have close decay widths. This is the reflection of chiral symmetry. For $B_c^{\ast-}$, both the $b\to c(u)$ and $\bar c\to \bar d(\bar s)$ are calculated. The channel $B_c^{\ast-}\to \bar D^0l^-\bar\nu_l$ is much smaller than those of other channels, the reason of which is that the CKM matrix element in this case is $V_{ub}=4.13\times 10^{-3}$ which is much smaller. With the total width estimated in Eq.~(6), the branching ratios of these channels are presented in the third column. For the decay channels of $B^{\ast-}$ and $B_s^{\ast0}$, our results are little larger than those of Ref.~\cite{qc01}. There are two main reasons for this. First, the wave functions in Ref.~\cite{qc01} are solutions of a relativistic scalar harmonic oscillator potential, while we get the wave functions by solving the instantaneous BS equation with a Cornell-like potential. Second, in Ref.~\cite{qc01}, the form factors at $Q^2\equiv(P-P_f)^2=0$ are calculated, and the explicit expressions are achieved by using the assumption of the pole structure. While in our calculation, the numerical results of the form factors at all the physical-allowed $Q^2$ can be achieved by applying Eq.~(2). For $B_c^{\ast-}$, the $\eta_cl^-\bar\nu_l$ channels were studied in Ref.~\cite{wzg14} by using the QCD sum rules. There the authors got the partial widths $6.86\times 10^{-15}$ GeV, $6.84\times 10^{-15}$ GeV, and $2.15\times 10^{-15}$ GeV for $l^-=e^-,~\mu^-$, and $\tau^-$, respectively, which are close to ours. The largest branching ratio comes from the channel $B_s^0e^-\bar\nu_e$, which is the order of $10^{-7}$. The partial widths of the $V\to V$ channels are presented in Table II. Compared with the $V\to P$ case, the results are $2\sim 3$ times larger. The branching ratios are also calculated, which shows the largest order of magnitude can reach $10^{-6}$.

\begin{figure}[ht]
\centering
\subfigure[$B^\ast\rightarrow D^{(\ast)}e\nu$]{\includegraphics[scale=0.3]{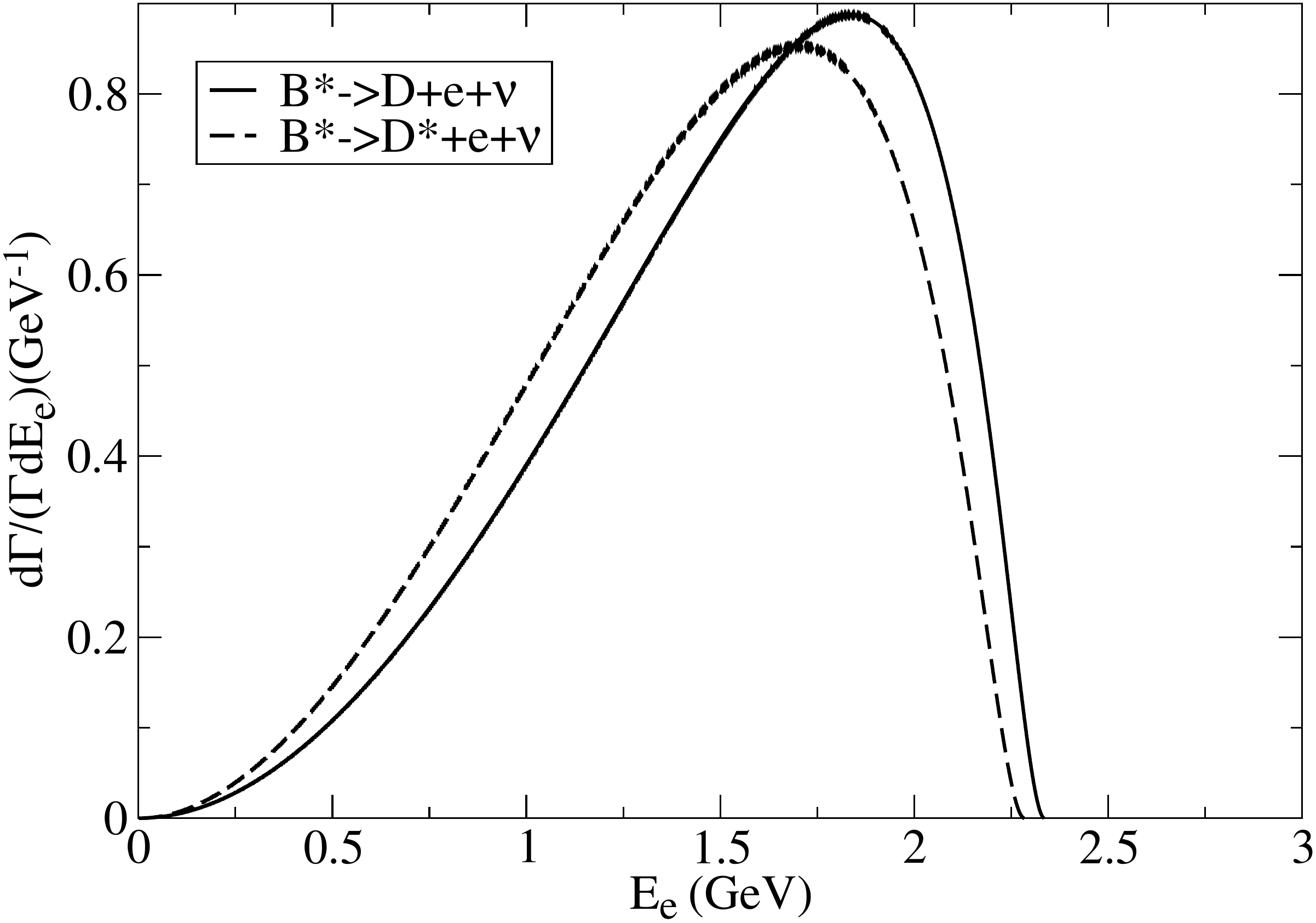}}
\hspace{1cm}
\subfigure[$B^\ast\rightarrow D^{(\ast)}\tau\nu$]{\includegraphics[scale=0.3]{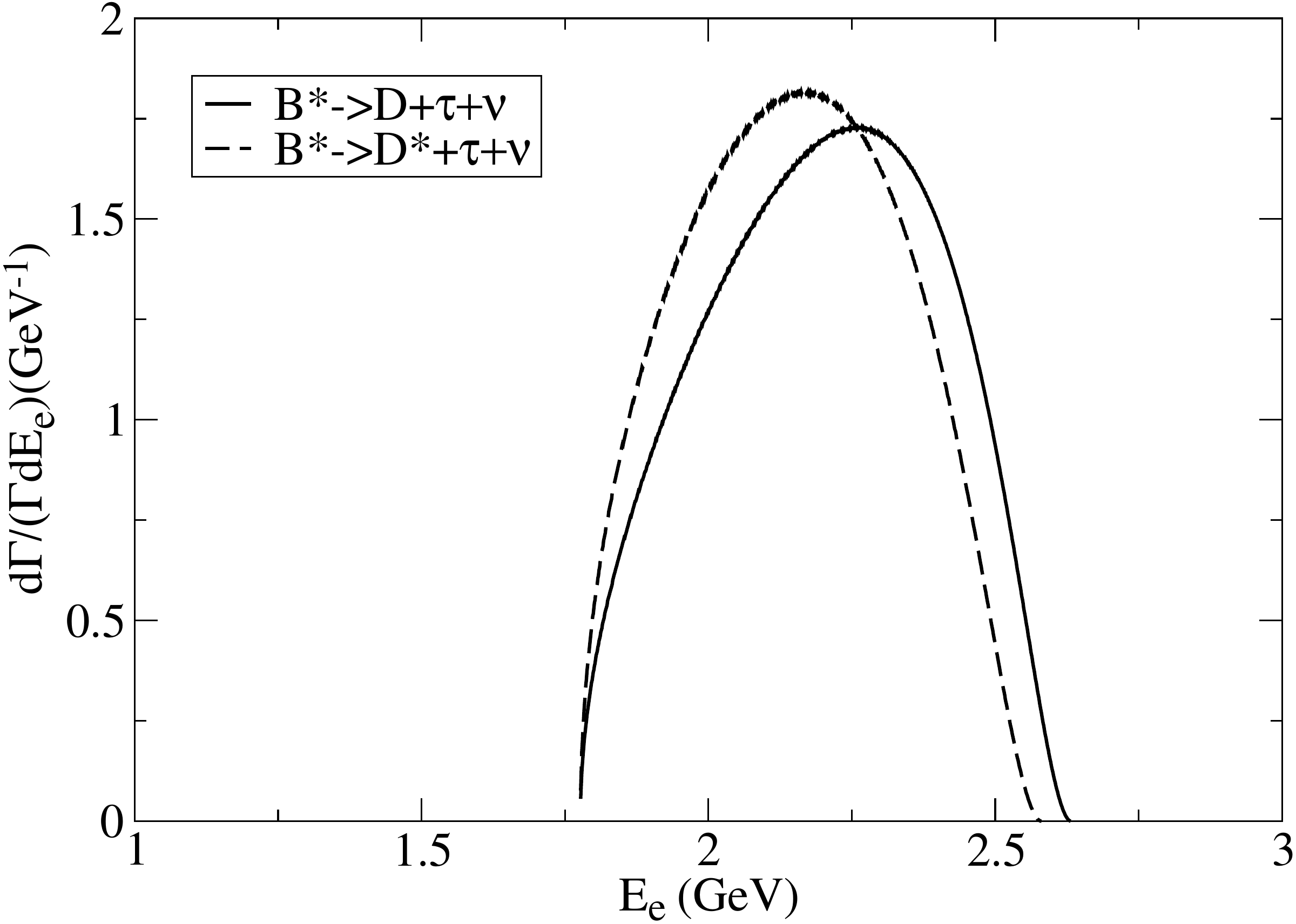}}
\caption[]{The energy spectra of final charged lepton in the $B^\ast\to D^{(\ast)}$ processes.}
\end{figure}

The energy spectra of the final charged lepton are presented in Figure $2\sim6$. For comparison, the results of $V\to P$ and $V\to V$ with the same final charged lepton are plotted in the same figure. For example, in Figure 2, the spectra of $B^{\ast-}\to D^{0}l^-\bar\nu_l$ and $B^{\ast-}\to D^{\ast0}l^-\bar\nu_l$ are presented. One can see that for $l^-=e^-$, when $E_l$ less (more) than about 1.6 GeV,  the spectrum of the pseudoscalar case is smaller (larger) than that of the vector case, and the peak value of the former is larger than that of the later. For $l^-=\tau^-$, the dividing point is at $E_\tau\simeq$ 2.25 GeV and the result for the peak value is reversed. This property is also owned by the $B_s^{\ast0}\to D_s^{(\ast)+}l^-\bar\nu_l$ (Figure 3) and $B_c^{\ast-}\to \eta_c(J/\psi)l^-\bar\nu_l$ (Figure 4) channels. The spectra of these three cases are quite similar to each other. The reason for this is that these decay channels have close phase space, which can be estimated by the mass difference of initial and final mesons: $M(B^{\ast-})-M(D^{(\ast)0})\simeq M(B_s^{\ast0})-M(D_s^{(\ast)+})\simeq M(B_c^{\ast-})-M(\eta_c(J/\psi))$. For the $B_c^{\ast-}\to \bar D^{(\ast)}l^-\bar\nu_l$ channels, $M(B_c^{\ast-})-M(\bar D^{(\ast)0})$ is more than 1 GeV larger than the former three cases, which makes the spectra (see Figure 5) have different forms. And the peak value for the $\bar D^0\tau^-\bar\nu_\tau$ channel gets larger than that of the $\bar D^{\ast0}\tau^-\bar\nu_\tau$ channel. For the $B_c^{\ast-}\to B_{d,s}^{(\ast)}e^-\bar\nu_e$ cases (Figure 6), when $E_e$ is around 0.45 GeV, the spectra reach the  maximum, which is larger for the pseudoscalar channels.

\begin{figure}[ht]
\centering
\subfigure[$B_s^\ast\rightarrow D_s^{(\ast)}e\nu$]{\includegraphics[scale=0.3]{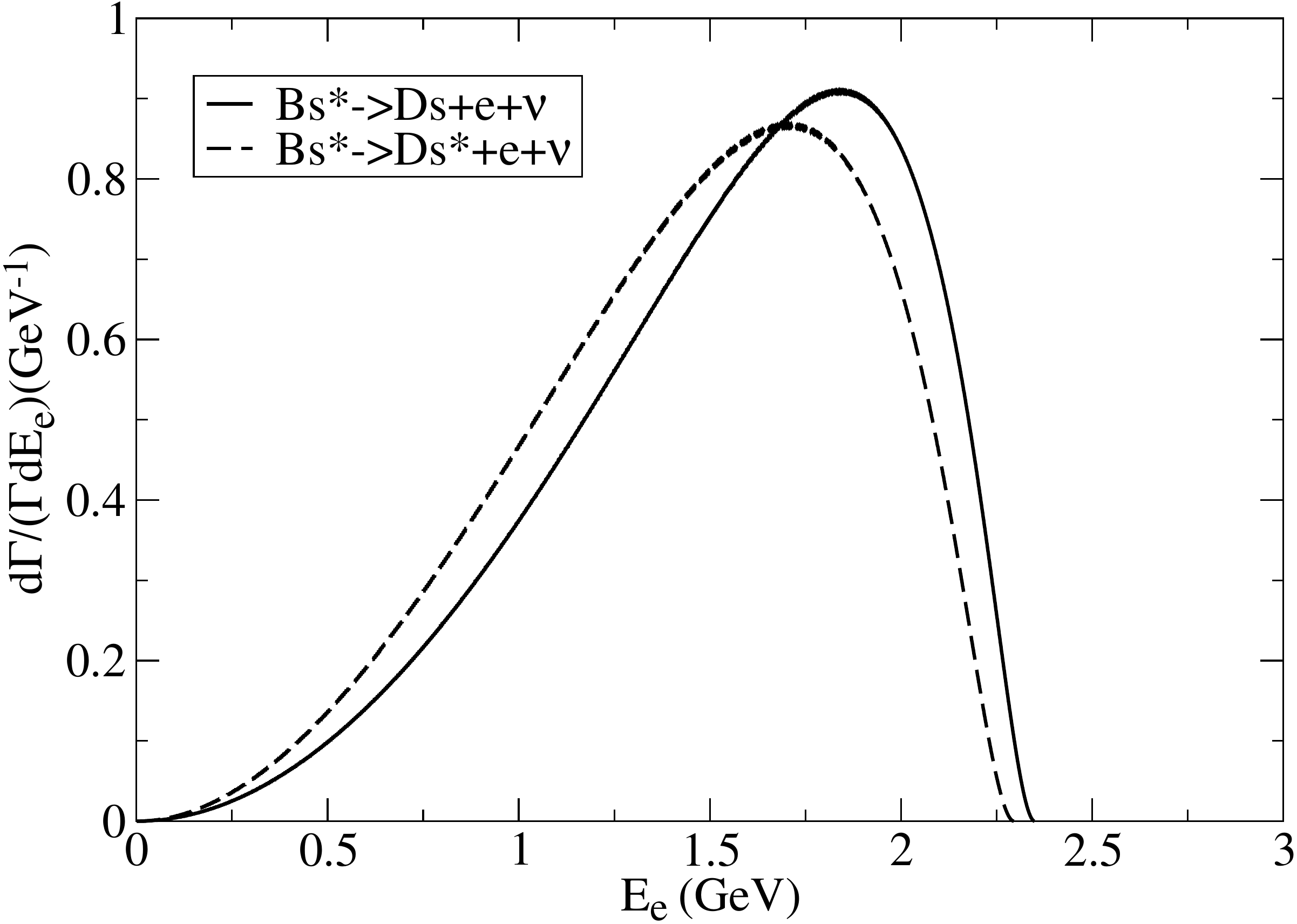}}
\hspace{1cm}
\subfigure[$B_s^\ast\rightarrow D_s^{(\ast)}\tau\nu$]{\includegraphics[scale=0.3]{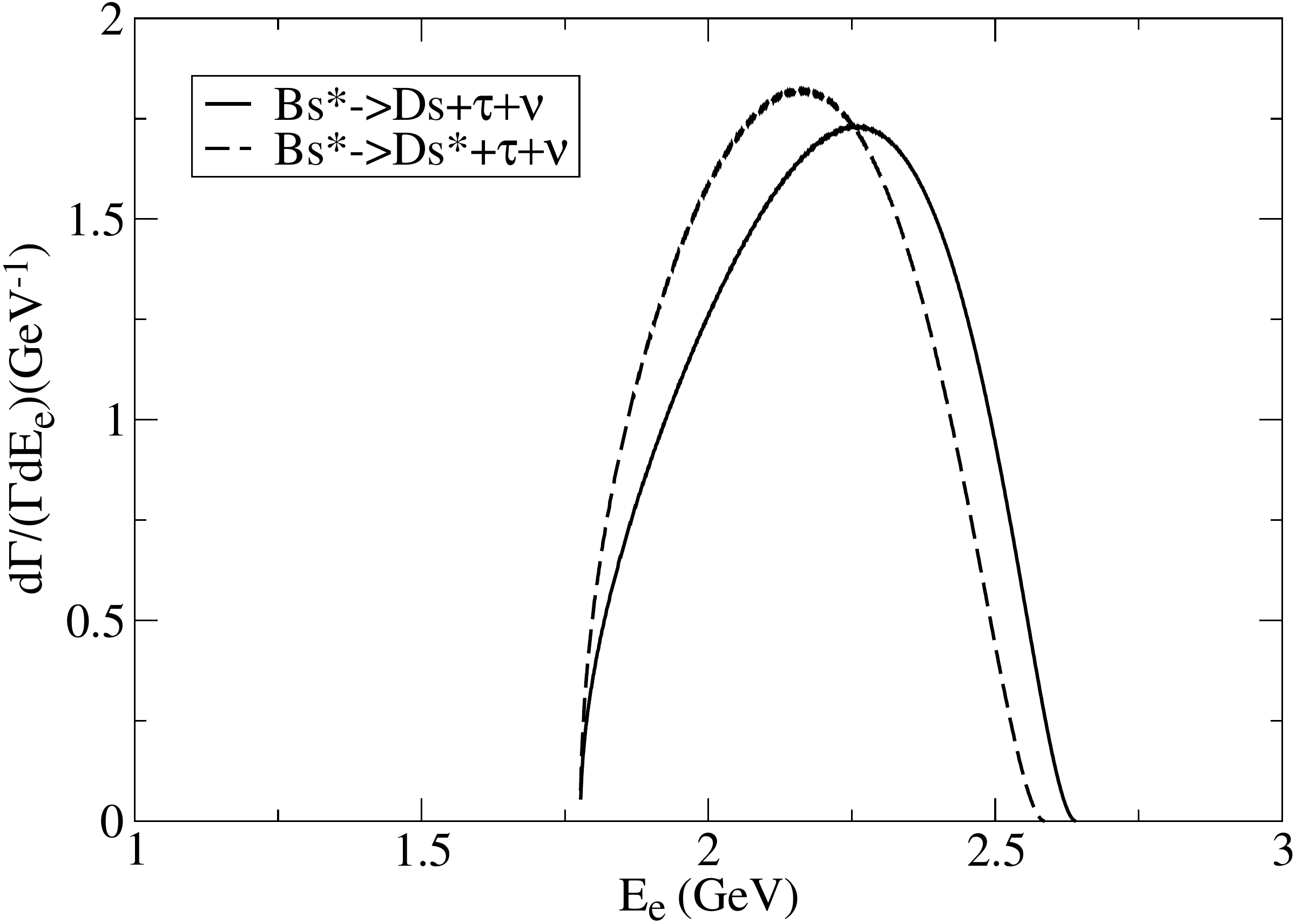}}
\caption[]{The energy spectra of final charged lepton in the $B_s^\ast\to D_s^{(\ast)}$ processes.}
\end{figure}

The ratio of the branching fractions is an interested quantity in experiments. Recently, the experimental results of this value for $B$, $B_s$, and $B_c$ sates have attracted more attentions as they deviate from the SM predictions by several standard deviations~\cite{LHCb15} (although the latest results from Belle~\cite{Belle17} are consistent with the SM prediction), which may indicate possible new physics beyond the SM~\cite{svj12}. If this is confirmed, similar results should also exist in their vector partners. In Table III, we present the ratios of the branching fractions for the vector cases. We define the following quantities
\begin{equation}
\begin{aligned}
\mathcal R=\frac{Br(V\to P \tau\bar\nu_\tau)}{Br(V\to Pe\bar\nu_e)},~~~~~~\mathcal R^\ast=\frac{Br(V\to V \tau\bar\nu_\tau)}{Br(V\to V e\bar\nu_e)}.
\end{aligned}
\end{equation}
One can see that for $B^{\ast-}\to D^0 (D^{\ast0})$, $B_s^{\ast0}\to D_s^+ (D_s^{\ast+})$, and $B_c^{\ast-}\to\eta_c (J/\psi)$, the results of $\mathcal R$($\mathcal R^\ast$) are close to each other. This is also the reflection of similar phase space. Besides that, one also notices that $\mathcal R$ is larger than $\mathcal R^\ast$ for these channels. For $B_c^{\ast-}\to \bar D^0 (\bar D^{\ast0})$, $\mathcal R$($\mathcal R^\ast$) is $2\sim 3$ times larger. The reason for this is that these channels have larger phase space. Also the relation between $\mathcal R$ and $\mathcal R^\ast$ reversed compared with former three cases. As the numerator and denominator in Eq.~8 share the same CKM matrix elements and part of uncertainties of the form factors which are canceled in the calculation, the two ratios are less model dependent and more robust, and can be compared with the future experimental results.

\begin{figure}[ht]
\centering
\subfigure[$B_c^\ast\rightarrow \eta_c (J/\psi) e\nu$]{\includegraphics[scale=0.3]{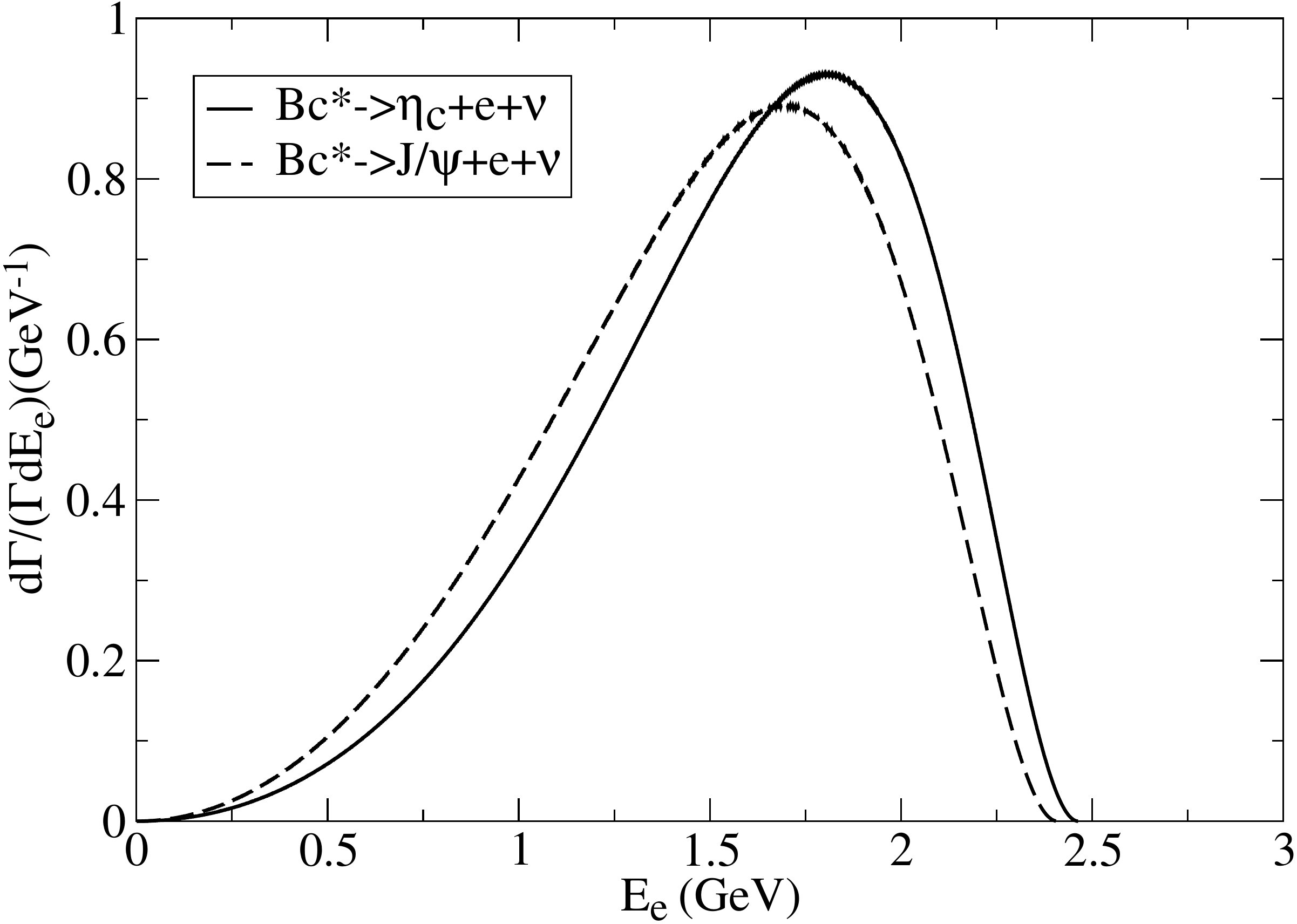}}
\hspace{1cm}
\subfigure[$B_c^\ast\rightarrow \eta_c (J/\psi) \tau\nu$]{\includegraphics[scale=0.3]{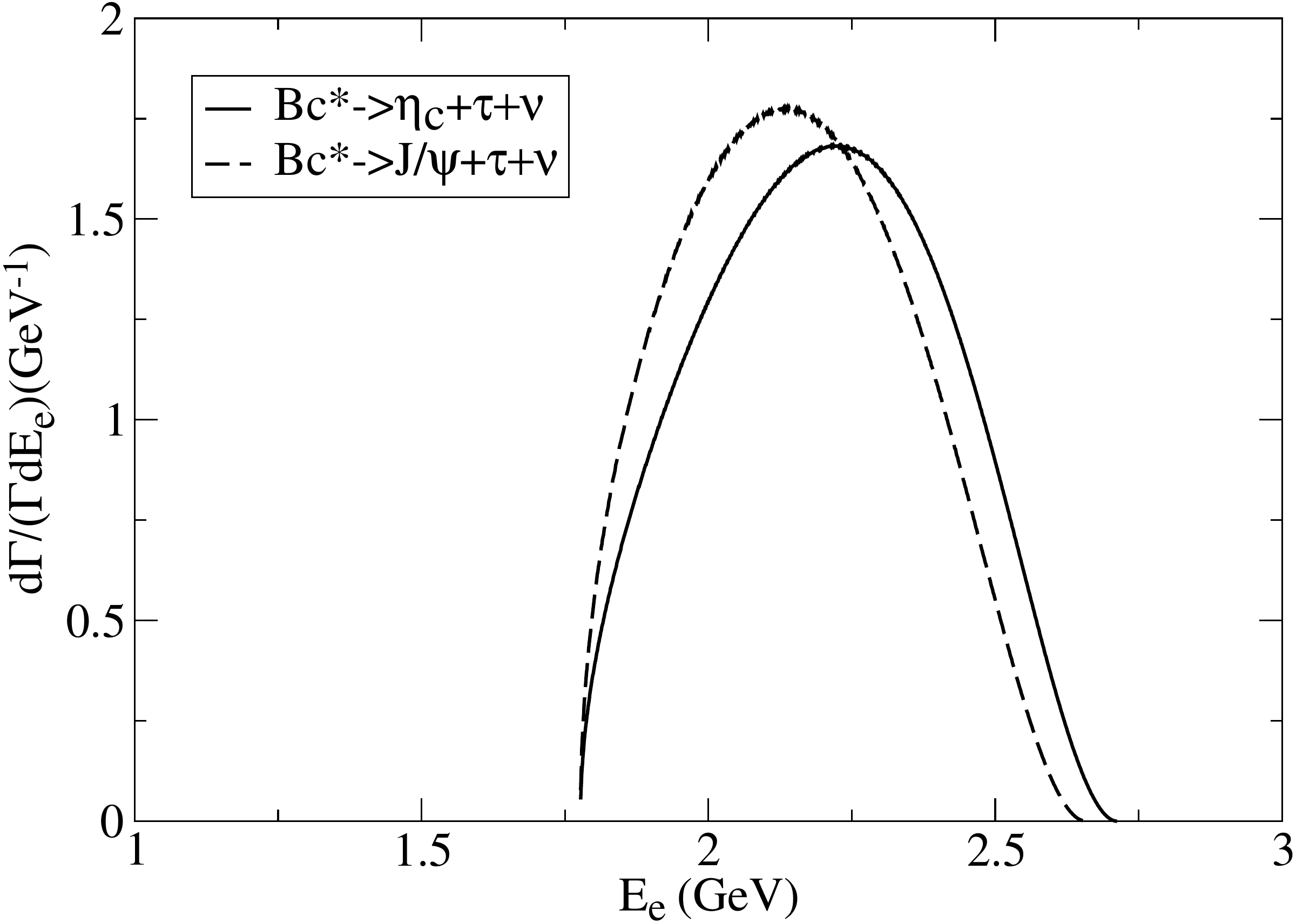}}
\caption[]{The energy spectra of final charged lepton in the $B_c^\ast\to\eta_c$ ($J\psi$) processes.}
\end{figure}

\begin{figure}[ht]
\centering
\subfigure[$B_c^\ast\rightarrow \bar D^{(\ast)} e\nu$]{\includegraphics[scale=0.30]{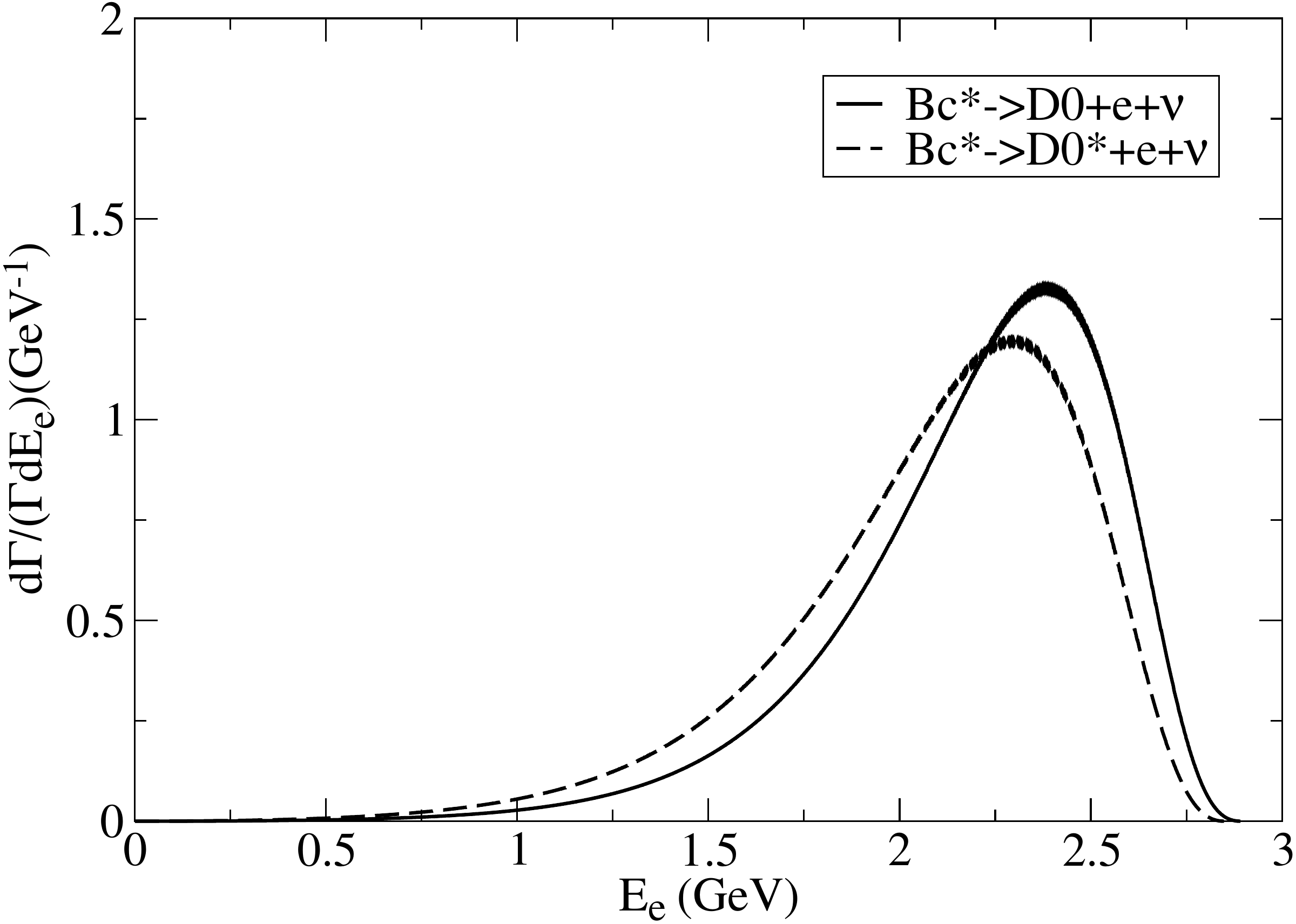}}
\hspace{1cm}
\subfigure[$B_c^\ast\rightarrow \bar D^{(\ast)} \tau\nu$]{\includegraphics[scale=0.30]{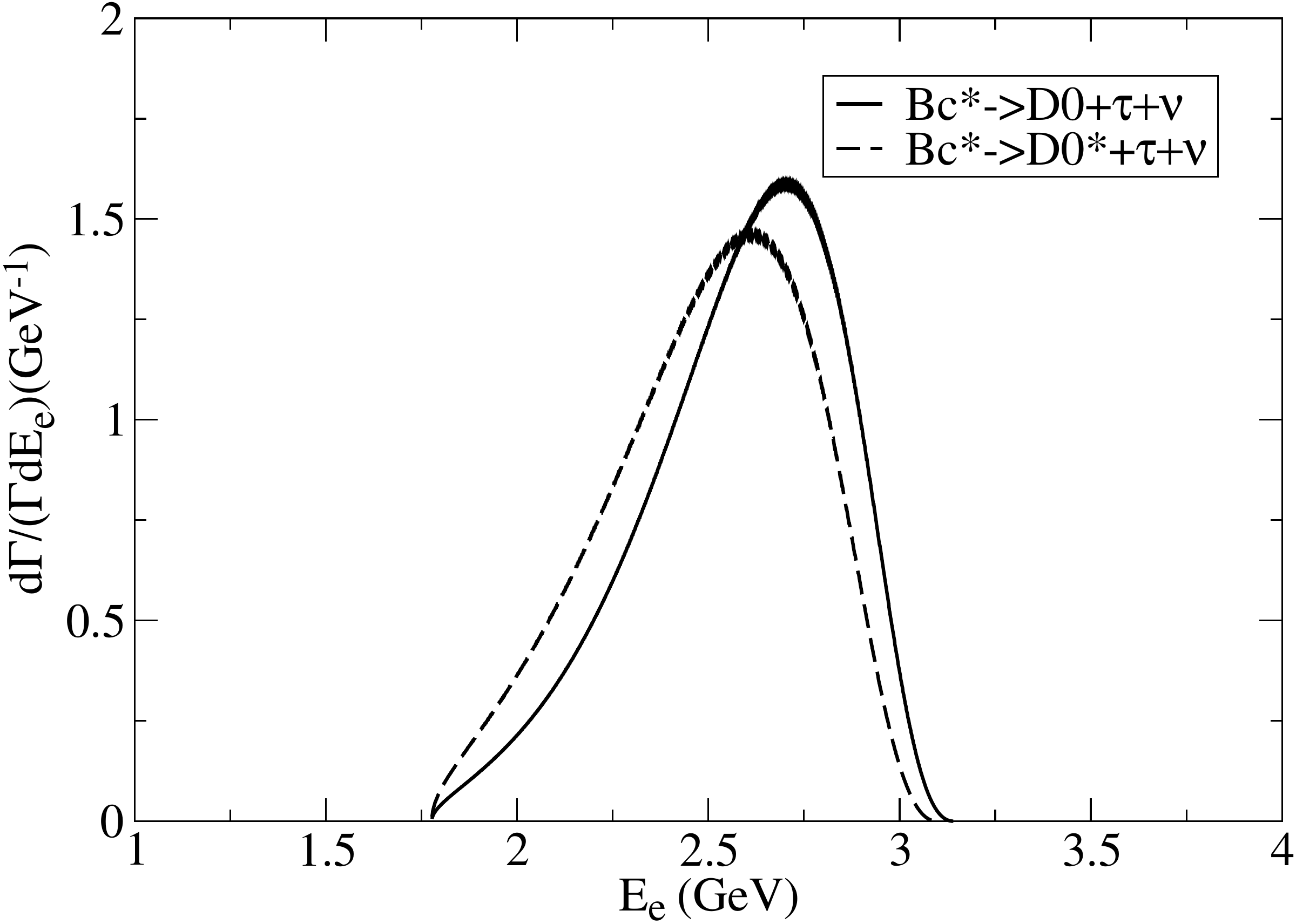}}
\caption[]{The energy spectra of final charged lepton in the $B_c^\ast\to\bar D^{(\ast)}$ processes.}
\end{figure}

\begin{figure}[ht]
\centering
\subfigure[$B_c^\ast\rightarrow B^{(\ast)} e\nu$]{\includegraphics[scale=0.3]{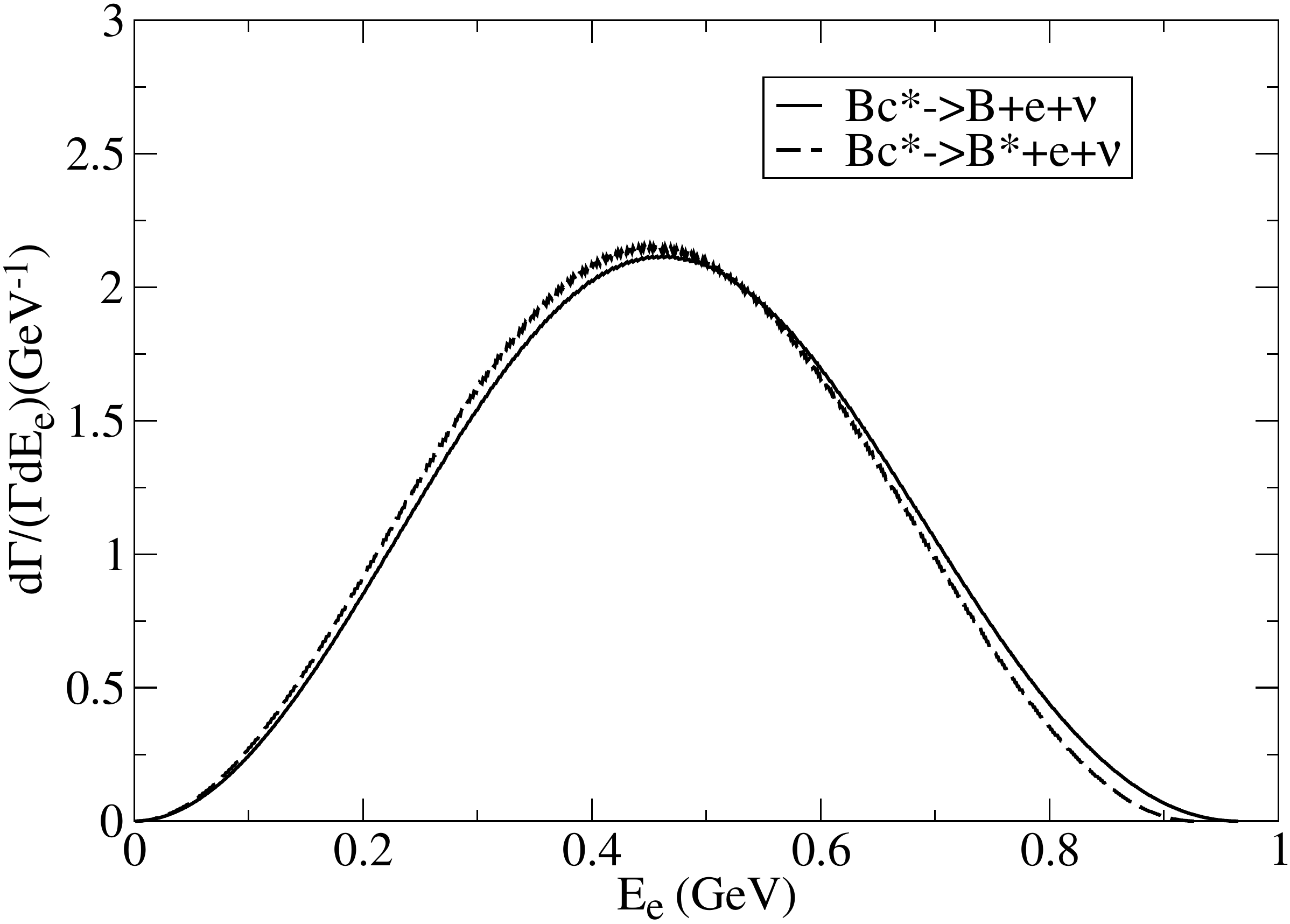}}
\hspace{1cm}
\subfigure[$B_c^\ast\rightarrow B_s^{(\ast)} e\nu$]{\includegraphics[scale=0.3]{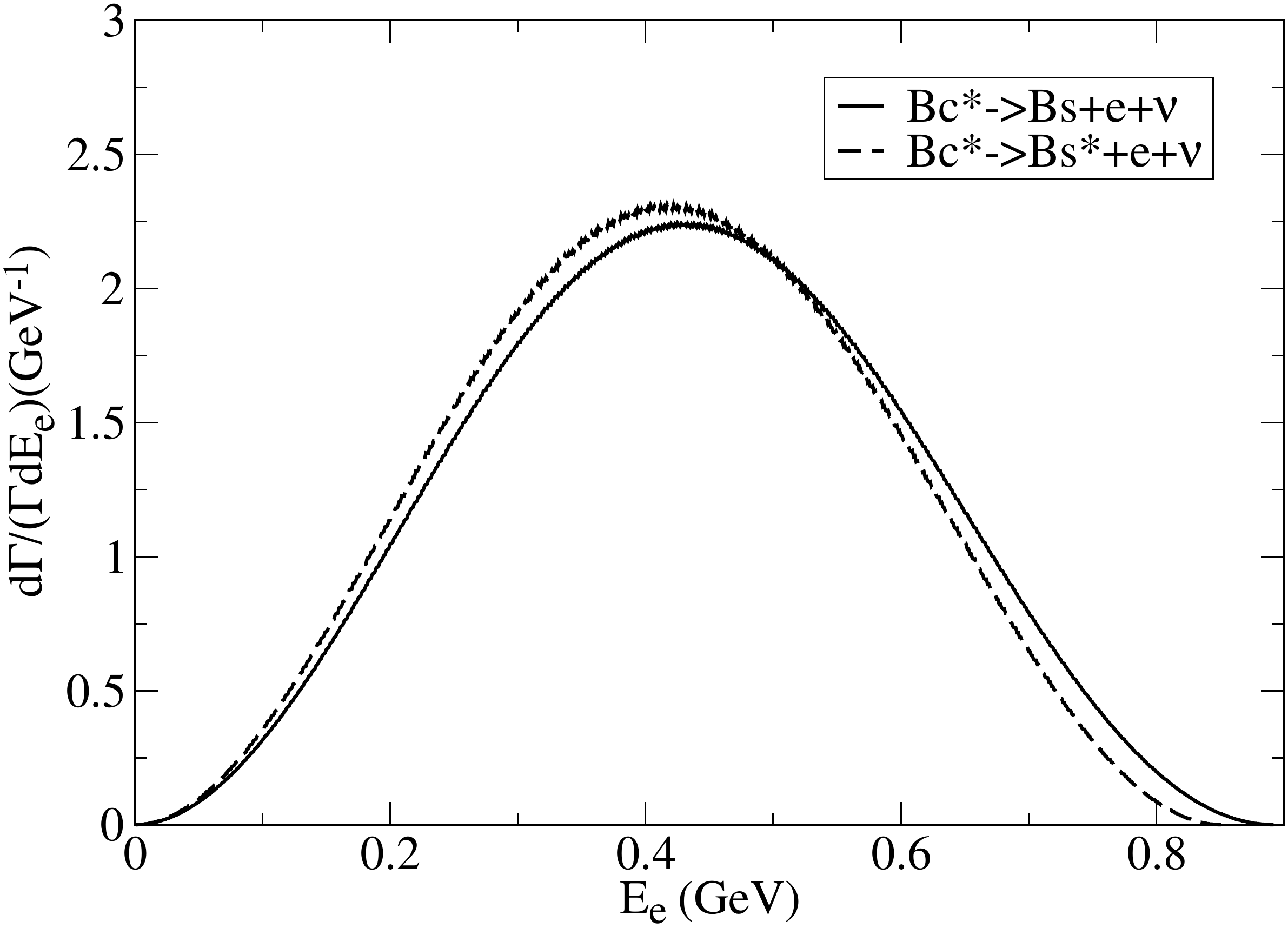}}
\caption[]{The energy spectra of final charged lepton in the $B_c^\ast\to B_{(d,s)}^{(\ast)}$ processes.}
\end{figure}

\begin{table}[htb]
 \caption{The partial decay widths (in units of GeV) and branching ratios of $B_{u,s,c}^\ast\to Pl^-\bar\nu_l$. The errors come from varying the parameters in our model by $\pm5\%$.}
 \label{semibb}
 \setlength{\tabcolsep}{0.15cm}
 \centering
\begin{tabular*}{\textwidth}{@{}@{\extracolsep{\fill}}ccccc}
\hline\hline
Channel &Width (This work)& Br (This work)&Br (Ref.~\cite{qc01})&Width (Ref.~\cite{wzg14}) \\ \hline
{\phantom{\Large{l}}}\raisebox{+.2cm}{\phantom{\Large{j}}}
$B^{\ast-}\rightarrow D^{0}e^{-}\bar\nu_{e}$  & $1.54^{+0.17}_{-0.15}\times 10^{-14}$&$3.29^{+0.36}_{-0.31}\times 10^{-8}$&$2.29\times 10^{-8}$&\\
{\phantom{\Large{l}}}\raisebox{+.2cm}{\phantom{\Large{j}}}
$B^{\ast-}\rightarrow D^{0}\mu^{-}\bar\nu_{\mu}$  &$1.53^{+0.17}_{-0.15}\times 10^{-14}$&$3.27^{+0.36}_{-0.31}\times 10^{-8}$&$2.29\times 10^{-8}$& \\
{\phantom{\Large{l}}}\raisebox{+.2cm}{\phantom{\Large{j}}}
$B^{\ast-}\rightarrow D^{0}\tau^{-}\bar\nu_{\tau}$  & $3.84^{+0.44}_{-0.38}\times 10^{-15}$&$8.21^{+0.93}_{-0.80}\times 10^{-9}$&$6.83\times 10^{-9}$&\\
{\phantom{\Large{l}}}\raisebox{+.2cm}{\phantom{\Large{j}}}
$B_s^{\ast0}\rightarrow D_{s}^{+}e^{-}\bar\nu_{e}$  & $1.39^{+0.15}_{-0.13}\times 10^{-14}$&$2.04^{+0.22}_{-0.19}\times 10^{-7}$&$1.39\times 10^{-7}$&\\
{\phantom{\Large{l}}}\raisebox{+.2cm}{\phantom{\Large{j}}}
$B_s^{\ast0}\rightarrow D_{s}^{+}\mu^{-}\bar\nu_{\mu}$  & $1.38^{+0.15}_{-0.13}\times 10^{-14}$&$2.03^{+0.22}_{-0.19}\times 10^{-7}$&$1.39\times 10^{-7}$&\\
{\phantom{\Large{l}}}\raisebox{+.2cm}{\phantom{\Large{j}}}
$B_s^{\ast0}\rightarrow D_{s}^{+}\tau^{-}\bar\nu_{\tau}$  & $3.64^{+0.42}_{-0.36}\times 10^{-15}$&$5.35^{+0.62}_{-0.53}\times 10^{-8}$&$4.08\times 10^{-8}$&\\
{\phantom{\Large{l}}}\raisebox{+.2cm}{\phantom{\Large{j}}}
$B_c^{\ast-}\rightarrow \eta_ce^{-}\bar\nu_{e}$ &$9.66^{+0.94}_{-0.84}\times 10^{-15}$&$4.20^{+0.41}_{-0.37}\times 10^{-7}$&&$6.86\times 10^{-15}$\\
{\phantom{\Large{l}}}\raisebox{+.2cm}{\phantom{\Large{j}}}
$B_c^{\ast-}\rightarrow \eta_c\mu^{-}\bar\nu_{\mu}$ &$9.63^{+0.94}_{-0.84}\times 10^{-15}$&$4.19^{+0.41}_{-0.37}\times 10^{-7}$&&$6.84\times 10^{-15}$\\
{\phantom{\Large{l}}}\raisebox{+.2cm}{\phantom{\Large{j}}}
$B_c^{\ast-}\rightarrow \eta_c\tau^{-}\bar\nu_{\tau}$ &$2.90^{+0.29}_{-0.26}\times 10^{-15}$&$1.26^{+0.13}_{-0.11}\times 10^{-7}$&&$2.15\times 10^{-15}$\\
{\phantom{\Large{l}}}\raisebox{+.2cm}{\phantom{\Large{j}}}
$B_c^{\ast-}\rightarrow \bar D^{0}e^{-}\bar\nu_{e}$ &$3.68^{+0.71}_{-0.60}\times 10^{-17}$&$1.60^{+0.31}_{-0.26}\times 10^{-9}$&&\\
{\phantom{\Large{l}}}\raisebox{+.2cm}{\phantom{\Large{j}}}
$B_c^{\ast-}\rightarrow \bar D^{0}\mu^{-}\bar\nu_{\mu}$ &$3.67^{+0.71}_{-0.60}\times 10^{-17}$&$1.60^{+0.31}_{-0.26}\times 10^{-9}$&&\\
{\phantom{\Large{l}}}\raisebox{+.2cm}{\phantom{\Large{j}}}
$B_c^{\ast-}\rightarrow \bar D^{0}\tau^{-}\bar\nu_{\tau}$ &$2.49^{+0.46}_{-0.40}\times 10^{-17}$&$1.08^{+0.20}_{-0.17}\times 10^{-9}$&&\\
{\phantom{\Large{l}}}\raisebox{+.2cm}{\phantom{\Large{j}}}
$B_c^{\ast-}\rightarrow B^{0}e^{-}\bar\nu_{e}$ &$1.33^{+0.19}_{-0.16}\times 10^{-15}$&$5.78^{+0.81}_{-0.71}\times 10^{-8}$&&\\
{\phantom{\Large{l}}}\raisebox{+.2cm}{\phantom{\Large{j}}}
$B_c^{\ast-}\rightarrow B^{0}\mu^{-}\bar\nu_{\mu}$ &$1.28^{+0.18}_{-0.16}\times 10^{-15}$&$5.57^{+0.78}_{-0.68}\times 10^{-8}$&&\\
{\phantom{\Large{l}}}\raisebox{+.2cm}{\phantom{\Large{j}}}
$B_c^{\ast-}\rightarrow B_s^{0}e^{-}\bar\nu_{e}$ &$2.17^{+0.29}_{-0.26}\times 10^{-14}$&$9.43^{+1.25}_{-1.11}\times 10^{-7}$&&\\
{\phantom{\Large{l}}}\raisebox{+.2cm}{\phantom{\Large{j}}}
$B_c^{\ast-}\rightarrow B_s^{0}\mu^{-}\bar\nu_{\mu}$ &$2.06^{+0.27}_{-0.24}\times 10^{-14}$&$8.96^{+1.19}_{-1.05}\times 10^{-7}$&&\\
\hline\hline
\end{tabular*}
\end{table}

\begin{table}[htb]
 \caption{The partial decay widths (in units of GeV) and branching ratios of $B_{u,s,c}^\ast\to Vl^-\bar\nu_l$. The errors come from varying the parameters in our model by $\pm5\%$.}
 \label{semibb}
 \setlength{\tabcolsep}{0.5cm}
 \centering
\begin{tabular*}{\textwidth}{@{}@{\extracolsep{\fill}}ccc}
\hline\hline
Channel&Width (GeV) & Br \\ \hline
{\phantom{\Large{l}}}\raisebox{+.2cm}{\phantom{\Large{j}}}
$B^{\ast-}\rightarrow D^{\ast0}e^{-}\bar\nu_{e}$  & $4.40^{+0.49}_{-0.42}\times 10^{-14}$ &$9.40^{+1.05}_{-0.89}\times 10^{-8}$ \\
{\phantom{\Large{l}}}\raisebox{+.2cm}{\phantom{\Large{j}}}
$B^{\ast-}\rightarrow D^{\ast0}\mu^{-}\bar\nu_{\mu}$  & $4.38^{+0.49}_{-0.41}\times 10^{-14}$&$9.36^{+1.04}_{-0.88}\times 10^{-8}$\\
{\phantom{\Large{l}}}\raisebox{+.2cm}{\phantom{\Large{j}}}
$B^{\ast-}\rightarrow D^{\ast0}\tau^{-}\bar\nu_{\tau}$  &$9.51^{+1.09}_{-0.92}\times 10^{-15}$ &$2.03^{+0.23}_{-0.20}\times 10^{-8}$\\
{\phantom{\Large{l}}}\raisebox{+.2cm}{\phantom{\Large{j}}}
$B_s^{\ast0}\rightarrow D_{s}^{\ast+}e^{-}\bar\nu_{e}$  &$3.89^{+0.42}_{-0.36}\times 10^{-14}$ &$5.72^{+0.62}_{-0.53}\times 10^{-7}$\\
{\phantom{\Large{l}}}\raisebox{+.2cm}{\phantom{\Large{j}}}
$B_s^{\ast0}\rightarrow D_{s}^{\ast+}\mu^{-}\bar\nu_{\mu}$  & $3.87^{+0.42}_{-0.36}\times 10^{-14}$&$5.69^{+0.62}_{-0.53}\times 10^{-7}$\\
{\phantom{\Large{l}}}\raisebox{+.2cm}{\phantom{\Large{j}}}
$B_s^{\ast0}\rightarrow D_{s}^{\ast+}\tau^{-}\bar\nu_{\tau}$  &$8.87^{+1.03}_{-0.87}\times 10^{-15}$ &$1.30^{+0.15}_{-0.13}\times 10^{-7}$\\
{\phantom{\Large{l}}}\raisebox{+.2cm}{\phantom{\Large{j}}}
$B_c^{\ast-}\rightarrow J/\psi e^{-}\bar\nu_{e}$ &$2.61^{+0.26}_{-0.23}\times 10^{-14}$&$1.13^{+0.11}_{-0.10}\times 10^{-6}$\\
{\phantom{\Large{l}}}\raisebox{+.2cm}{\phantom{\Large{j}}}
$B_c^{\ast-}\rightarrow J/\psi\mu^{-}\bar\nu_{\mu}$ &$2.60^{+0.26}_{-0.23}\times 10^{-14}$&$1.13^{+0.11}_{-0.10}\times 10^{-6}$\\
{\phantom{\Large{l}}}\raisebox{+.2cm}{\phantom{\Large{j}}}
$B_c^{\ast-}\rightarrow J/\psi\tau^{-}\bar\nu_{\tau}$ &$7.21^{+0.75}_{-0.65}\times 10^{-15}$&$3.13^{+0.32}_{-0.28}\times 10^{-7}$\\
{\phantom{\Large{l}}}\raisebox{+.2cm}{\phantom{\Large{j}}}
$B_c^{\ast-}\rightarrow \bar D^{\ast0}e^{-}\bar\nu_{e}$ &$1.06^{+0.21}_{-0.17}\times 10^{-16}$&$4.61^{+0.92}_{-0.76}\times 10^{-9}$\\
{\phantom{\Large{l}}}\raisebox{+.2cm}{\phantom{\Large{j}}}
$B_c^{\ast-}\rightarrow \bar D^{\ast0}\mu^{-}\bar\nu_{\mu}$ &$1.06^{+0.21}_{-0.17}\times 10^{-16}$&$4.61^{+0.92}_{-0.76}\times 10^{-9}$\\
{\phantom{\Large{l}}}\raisebox{+.2cm}{\phantom{\Large{j}}}
$B_c^{\ast-}\rightarrow \bar D^{\ast0}\tau^{-}\bar\nu_{\tau}$ &$7.41^{+1.40}_{-1.16}\times 10^{-17}$&$3.22^{+0.61}_{-0.51}\times 10^{-9}$\\
{\phantom{\Large{l}}}\raisebox{+.2cm}{\phantom{\Large{j}}}
$B_c^{\ast-}\rightarrow B^{\ast0}e^{-}\bar\nu_{e}$ &$3.66^{+0.49}_{-0.48}\times 10^{-15}$&$1.59^{+0.21}_{-0.21}\times 10^{-7}$\\
{\phantom{\Large{l}}}\raisebox{+.2cm}{\phantom{\Large{j}}}
$B_c^{\ast-}\rightarrow B^{\ast0}\mu^{-}\bar\nu_{\mu}$ &$3.51^{+0.51}_{-0.43}\times 10^{-15}$&$1.53^{+0.22}_{-0.19}\times 10^{-7}$\\
{\phantom{\Large{l}}}\raisebox{+.2cm}{\phantom{\Large{j}}}
$B_c^{\ast-}\rightarrow B_s^{\ast0}e^{-}\bar\nu_{e}$ &$5.68^{+0.76}_{-0.67}\times 10^{-14}$&$2.47^{+0.33}_{-0.29}\times 10^{-6}$\\
{\phantom{\Large{l}}}\raisebox{+.2cm}{\phantom{\Large{j}}}
$B_c^{\ast-}\rightarrow B_s^{\ast0}\mu^{-}\bar\nu_{\mu}$ &$5.39^{+0.72}_{-0.63}\times 10^{-14}$&$2.34^{+0.31}_{-0.28}\times 10^{-6}$\\
\hline\hline
\end{tabular*}
\end{table}

\begin{table}[htb]
 \caption{The ratios of branching fractions of different decay channels of $B_{u,s,c}^\ast$.}
 \label{semibb}
 \setlength{\tabcolsep}{0.5cm}
 \centering
\begin{tabular*}{\textwidth}{@{}@{\extracolsep{\fill}}cccc}
\hline\hline
Channel&$\mathcal R$ &Channel& $\mathcal R^\ast$ \\ \hline
{\phantom{\Large{l}}}\raisebox{+.2cm}{\phantom{\Large{j}}}
$B^{\ast-}\rightarrow D^{0}$  &0.249 &$B^{\ast-}\rightarrow D^{\ast0}$ &0.216 \\
{\phantom{\Large{l}}}\raisebox{+.2cm}{\phantom{\Large{j}}}
$B_s^{\ast0}\rightarrow D_{s}^{+}$  & 0.262&$B_s^{\ast0}\rightarrow D_{s}^{\ast+}$ &0.228\\
{\phantom{\Large{l}}}\raisebox{+.2cm}{\phantom{\Large{j}}}
$B_c^{\ast-}\rightarrow \eta_c$ &0.300&$B_c^{\ast-}\rightarrow J/\psi$&0.276\\
{\phantom{\Large{l}}}\raisebox{+.2cm}{\phantom{\Large{j}}}
$B_c^{\ast-}\rightarrow \bar D^{0}$ &0.677&$B_c^{\ast-}\rightarrow \bar D^{\ast0}$&0.699\\
\hline\hline
\end{tabular*}
\end{table}

\section{Conclusions}
As a conclusion, we have studied the semileptonic decays of the $b$-flavored vector heavy mesons. Both cases for the final meson being a pseudoscalar or vector are considered. The partial widths of these channels are of the order of $10^{-14}\sim10^{-17}$ GeV. As the single-photon decay channel is dominant, its partial width is used to estimate the total width of the initial meson. As a result, for $B^{\ast-}$, the $D^{\ast0}e^-\bar\nu_e$ channel has the largest branching ratio $9.40\times10^{-8}$; for $B_s^{\ast0}$, the $D_s^{\ast+}e^-\bar\nu_e$ channel has the largest branching ratio $5.72\times10^{-7}$; for $B_c^{\ast-}$, the $B_s^{\ast0}e^-\bar\nu_e$ channel has the largest branching ratio $2.47\times10^{-6}$. Experimental results for these channels at LHCb and future B-factories are expected, which will be helpful to set more stringent constraint on the SM parameters and clarify the possible anomalies observed in the semileptonic decays of $b$-flavored pseudoscalar mesons.

\section{Acknowledgments}

This paper was supported in part by the National Natural Science
Foundation of China (NSFC) under Grant No.~11405037, No.~11505039 and No.~11575048.


 \begin{appendices}
 \section*{Appendix}
 
The quantity $\varphi$ is constructed with momenta and gamma matrices by considering corresponding spin and parity properties. For the $1^-$ initial  state, it has the form 
\begin{equation}
\begin{aligned}
\varphi_{1^-}(q_\perp)&=(q_\perp\cdot\epsilon)\left[f_1(q_\perp)+\frac{\slashed{P}}{M}f_2(q_\perp)
+\frac{\slashed{q}_\perp}{M}f_3(q_\perp)+\frac{\slashed{P}\slashed{q}_\perp}{M^2}f_4(q_\perp)\right]\\
&~~~~~+ M\slashed\epsilon\left[f_5(q_\perp)+\frac{\slashed{P}}{M}f_6(q_\perp)
+\frac{\slashed{q}_\perp}{M}f_7(q_\perp)+\frac{\slashed{P}\slashed{q}_\perp}{M^2}f_8(q_\perp)\right],
\end{aligned}
\end{equation}
where $f_i$s are functions of $q_\perp^2$.
For the $0^-$ final  state, it can be written as
\begin{equation}
\begin{aligned}
\varphi_{0^-}(q_{f\perp})&=\left[g_1(q_{f\perp})+\frac{\slashed{P}_f}{M_f}g_2(q_{f\perp})
+\frac{\slashed{q}_{f\perp}}{M_f}g_3(q_{f\perp})+\frac{\slashed{P}_f\slashed{q}_{f\perp}}{M_f^2}g_4(q_{f\perp})\right]\gamma_5,
\end{aligned}
\end{equation}
where we have used $q^\mu_{f\perp}=q_f^\mu-\frac{P\cdot q_f}{M^2}P^\mu$; $g_i$s are functions of $q_{f\perp}^2$.
The numerical results of $f_i$ and $g_i$ can be achieved by solving Eq.~(2). In the calculation, not all the $f_i$s or $g_i$s are independent, as the last two equations in Eq.~(2) provide the constraint conditions. For the $1^-$ state, we choose $f_3$, $f_4$, $f_5$, $f_6$ as the independent variables, and for the $0^-$ state, we choose $g_1$ and $g_2$.

The positive energy part of $\varphi$ are kept in the calculation. For the $1^-$ state, it has the form
\begin{equation}
\begin{aligned}
\varphi^{++}_{1^-}(q_\perp)&=(q_\perp\cdot\epsilon)\left[A_1(q_\perp)+\frac{\slashed{P}}{M}A_2(q_\perp)
+\frac{\slashed{q}_\perp}{M}A_3(q_\perp)+\frac{\slashed{P}\slashed{q}_\perp}{M^2}A_4(q_\perp)\right]\\
&~~~~~+ M\slashed\epsilon\left[A_5(q_\perp)+\frac{\slashed{P}}{M}A_6(q_\perp)
+\frac{\slashed{q}_\perp}{M}A_7(q_\perp)+\frac{\slashed{P}\slashed{q}_\perp}{M^2}A_8(q_\perp)\right],
\end{aligned}
\end{equation}
where $A_i$s are defined as
\begin{equation}
\begin{aligned}
&A_{1}=\frac{(\omega_{1}+\omega_{2})q_{\perp}^{2}f_{3}+(m_{1}+m_{2})q_{\perp}^{2}f_{4}+2M^{2}\omega_{2}f_{5}-2M^{2}m_{2}f_{6}}{2M(m_{1}\omega_{2}+m_{2}\omega_{1})},\\
&A_{2}=\frac{(m_{1}-m_{2})q_{\perp}^{2}f_{3}+(\omega_{1}-\omega_{2})q_{\perp}^{2}f_{4}-2M^{2}m_{2}f_{5}+2M^{2}\omega_{2}f_{6}}{2M(m_{1}\omega_{2}+m_{2}\omega_{1})},\\
&A_{3}=\frac{1}{2}(f_{3}+\frac{m_{1}+m_{2}}{\omega_{1}+\omega_{2}}f_{4}-\frac{2M^{2}}{m_{1}\omega_{2}+m_{2}\omega_{1}}f_{6}),\\
&A_{4}=\frac{1}{2}(\frac{\omega_{1}+\omega_{2}}{m_{1}+m_{2}}f_{3}+f_{4}-\frac{2M^{2}}{m_{1}\omega_{2}+m_{2}\omega_{1}}f_{5}),\\
&A_{5}=\frac{1}{2}(f_{5}-\frac{\omega_{1}+\omega_{2}}{m_{1}+m_{2}}f_{6}),~~~~~A_{6}=\frac{1}{2}(-\frac{m_{1}+m_{2}}{\omega_{1}+\omega_{2}}f_{5}+f_{6}),\\
&A_{7}=A_{5}\frac{M(\omega_{1}-\omega_{2})}{m_{1}\omega_{2}+m_{2}\omega_{1}},~~~~~~~~A_{8}=A_{6}\frac{M(\omega_{1}+\omega_{2})}{m_{1}\omega_{2}+m_{2}\omega_{1}}.\\
\end{aligned}
\end{equation}

And for the $0^-$ final state, the positive energy part of $\varphi$ has the form
\begin{equation}
\begin{aligned}
\varphi^{++}_{0^-}(q_{f\perp})&=\left[B_1(q_{f\perp})+\frac{\slashed{P}_f}{M_f}B_2(q_{f\perp})
+\frac{\slashed{q}_{f\perp}}{M_f}B_3(q_{f\perp})+\frac{\slashed{P}_f\slashed{q}_{f\perp}}{M_f^2}B_4(q_{f\perp})\right]\gamma_5,
\end{aligned}
\end{equation}
where
\begin{equation}
\begin{aligned}
&B_{1}=\frac{M_{f}}{2}(\frac{\omega^\prime_{1}+\omega^\prime_{2}}{m^\prime_{1}+m^\prime_{2}}g_{1}+g_{2}),\\
&B_{2}=\frac{M_{f}}{2}(g_{1}+\frac{m^\prime_{1}+m^\prime_{2}}{\omega^\prime_{1}+\omega^\prime_{2}}g_{2}),\\
&B_{3}=-\frac{M_{f}(\omega^\prime_{1}-\omega^\prime_{2})}{m^\prime_{1}\omega^\prime_{2}+m^\prime_{2}\omega^\prime_{1}}B_{1},\\
&B_{4}=-\frac{M_{f}(\omega^\prime_{1}+\omega^\prime_{2})}{m^\prime_{1}\omega^\prime_{2}+m^\prime_{2}\omega^\prime_{1}}B_{2}.\\
\end{aligned}
\end{equation}
Here we have used the definitions $\omega^\prime_1=\sqrt{m^{\prime2}_1-q_{f\perp}^2}$ and $\omega^\prime_2=\sqrt{m^{\prime2}_2-q_{f\perp}^2}$, where $m_1^\prime$ and $m_2^\prime$ are respectively the masses of quark and antiquark in the final meson.

 \end{appendices}
 

\end{document}